\documentclass[epj]{svjour}
\usepackage{latexsym}
\usepackage{graphics}
\begin{document}
\title{Active and driven hydrodynamic crystals}
\author{Nicolas Desreumaux\inst{1} \and Nicolas Florent\inst{2}\and Eric Lauga\inst{2} \and Denis Bartolo\inst{1}}
\institute{Laboratoire de Physique et M\'ecanique des Milieux H\'et\'erog\'enes, CNRS, ESPCI, Universit\'e Paris 6, Universit\'e Paris 7,\\ 10, rue Vauquelin, 75005 Paris FRANCE. \and Department of Mechanical and Aerospace Engineering, 
University of California San Diego, 9500 Gilman Drive,\\  La Jolla CA 92093-0411, USA.}
\date{Received: date / Revised version: date}

\abstract{Motivated by the experimental ability to produce monodisperse particles in microfluidic devices, we study theoretically the hydrodynamic stability of driven and active crystals. We first recall the theoretical tools allowing to quantify the dynamics of  elongated particles in a confined fluid. In this regime hydrodynamic interactions between particles arise from a superposition of potential dipolar singularities. We exploit this feature to derive the equations of motion for the  particle positions and orientations. After showing that all five planar Bravais lattices are stationary solutions of the equations of motion, we consider separately the case where the particles are passively driven by an external force, and the situation where they are self-propelling. We first demonstrate that phonon modes propagate in  driven crystals, which are always marginally stable. The spatial structure of the eigenmodes depend solely on the symmetries of the lattices, and on the orientation of the driving force.  For active crystals, the stability of the particle positions and orientations depends not only on the symmetry of the crystals but also on the perturbation wavelengths and on the crystal density. Unlike unconfined fluids, the stability of active crystals is independent of the nature of the propulsion mechanism at the single particle level.   The square and rectangular lattices are found to be linearly unstable at short wavelengths provided the volume fraction of the crystals is high enough. Differently, hexagonal, oblique, and face-centered crystals are always unstable. Our work provides a theoretical basis for future experimental work on flowing microfluidic crystals.
}\maketitle

\section{Introduction}
\label{intro}

The dynamics of passive  suspensions is a field with a long history  in physical hydrodynamics. Much effort has been devoted  to understand, e.g., the origin of fluctuations in the sedimentation of spheres under gravity as well as instabilities in   suspensions of elongated fibers (see reviews in  \cite{Ramaswamy_sedimentation,Guazzelli:2011ec} and references therein).  More recently, a significant experimental \cite{WuLibchaber2000,Dombrowski2004,CisnerosExp2007}
 and theoretical \cite{BaskaranMarchetti2009,Ramaswamy:2010bf,Kochreview} research effort has focused on the dynamics of active suspensions where instead of having particles driven by an external field (e.g.~gravity), one considers the dynamics and interactions of self-propelled synthetic or biological swimmers. In this case, the interplay of activity and hydrodynamic interactions leads to long-wavelength instabilities \cite{Saintillan:2007jr,SaintillanShelley2008}.

Most of the past work on passive (driven) and active suspensions has focused on instabilities and fluctuating behavior in three-dimensional systems. However, over the last ten years microfluidics has offered a number of simple and effective solutions to produce and  manipulate large ensemble of highly monodisperse microparticles,  prone to form crystal structure in quasi-two dimensional channels \cite{Dendukuri:2009fpa}. For driven particles, these technological advances have motivated, for example, the study of the nonlinear dynamics of finite flowing crystals \cite{Baron:2008cp,Blawzdziewicz:2010dj}, phonons in one dimensional microfluic-droplet crystals
\cite{Beatus:2006hc} and flowing lattices of bubbles \cite{Hashimoto:2006wn}. In the case of active particles, these fabrication methods could be extended to self-propelled catalytic colloids \cite{Howse:2007ed,Paxton:2004we} or reactive droplets  \cite{Thutupalli:2011bv}.

Motivated by these advances,  we take in this paper an approach contrasting with the traditional study of disordered  suspensions and consider the dynamics of confined driven and active hydrodynamic crystals. We first develop a formalism to study theoretically position and orientation instabilities for flowing discrete suspensions under confinement. In the case of driven particles, we demonstrate formally that all crystals are marginally stable and study in detail the eigenmodes of deformation for all five two-dimensional Bravais lattices. For active particles, 
we show that  square and rectangular crystals are linearly unstable at short wavelengths provided the volume fraction of the crystals is high enough. Differently, hexagonal and oblique (resp.~face-centered) crystals are always unstable for long- (resp.~short-) wavelength perturbations. In contrast with past work on three-dimensional swimmers suspensions, the stability of confined active crystals is found to be independent of the pusher vs. puller nature of the actuation of individual active particles \cite{saintillan08}.

\section{Theoretical setup}
\subsection{Particle crystal in a confined fluid}
\begin{figure}\center
\resizebox{0.8\columnwidth}{!}{
\includegraphics{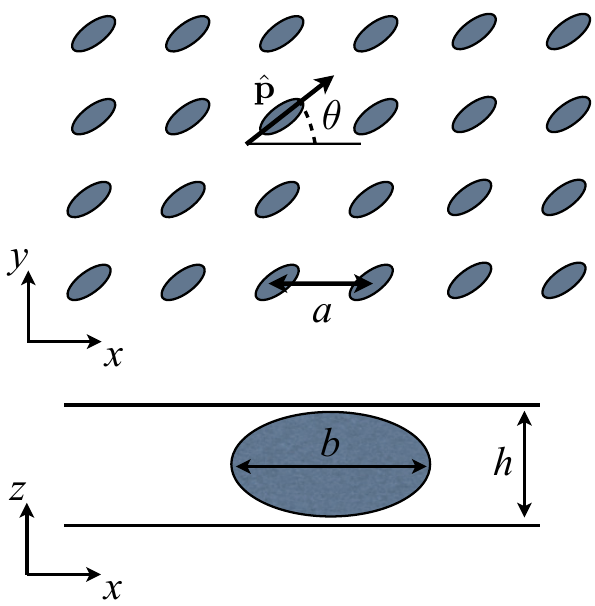}}
\caption{Schematic representation of the problem addressed in this paper: An extended hydrodynamic crystal is composed of anisotropic particles confined in a channel of height $h$ which are either actively swimming or passively driven by an external force (top- and side-views).}
\label{fig1}      
\end{figure}

We start by  describing the theoretical framework we use to quantify the large scale dynamics of  both active and driven microfluidic crystals. We focus our study on the case of identical particles living in quasi-bidimensional fluids, as sketched in Fig.~\ref{fig1}. The fluid is Newtonian and has a homogeneous thickness $h$ in the $z$-direction, comparable to the size of the particles. Our formalism will be valid both for thin films lying on a solid substrate (with one free surface and one no-slip wall), as well as  for microfluidic geometries where the fluid is confined between two parallel plates. The particles  can be either axisymmetric or anisotropic and are organized in two-dimensional crystal, see~Fig.~\ref{fig1}.  If $a$ denotes the typical lattice spacing of the crystal and $b$ the typical extent of the particle in the $(x,y)$ plane, we consider in this paper the dynamics in the dilute limit, e.g.~$a \gg b$.  In this limit, each particle $i$ is appropriately modeled as a pointwise body characterized by its in-plane position, ${\bf R}_i(t)\equiv(x_i(t),y_i(t))$, and its in-plane orientation, $\hat{\bf p}_i(t)$, where $\hat {\bf p}_i$ is a unit vector making an angle $\theta_i(t)$ with the $\hat {\bf x}$-axis. Having microscopic systems in mind, we  neglect the particle inertia and work in the limit of zero Reynolds number. In this Hele-Shaw setup, it is a classical result that the fluid flow is potential \cite{batchelor}.   The $z$-averaged fluid velocity, $\bf V$ and the $z$-averaged pressure, $P$,  are therefore related by
\begin{equation}
{\bf V(r)}=-G\nabla P,
\label{potential}
\end{equation}
where $G= \alpha h^2/\eta$; here $\eta$ is the fluid viscosity, and $\alpha=1/3$ for a thin film, and $\alpha=1/12$ for a shallow microchannel.
Together with  incompressibility, $\nabla\cdot\bf V=0$, Eq.~\ref{potential} determines the fluid flow and stress away from the particles.

We henceforth consider either swimmers moving  along their principal axis $\hat {\bf p}_i $, or passive particles driven by a uniform force field oriented along the $x$-direction (gravitational, electrostatic, magnetic,...). In all cases, the speed of an isolated particle in a quiescent fluid  is constant and denoted $U_0$. In addition to their individual dynamics, particles also follow the  surrounding flow, and  the equation of motion for particle $i$ thus reads 
\begin{equation}
\partial_t{\bf R}_i=U_0\hat {\bf q}+\mu{\bf V}({\bf R}_i),
\label{eq2}
\end{equation}
where $\hat {\bf q}=\hat {\bf p}_i$ for swimmers, $\hat {\bf q}=\hat {\bf x}$ for driven particles, and $\mu$ is a non-dimensional mobility coefficient~\cite{Beatus:2006hc,Beatus:2007ju}. Passive tracers have $\mu=1$. Conversely, for thick particles, the friction against the solid wall(s) can significantly reduce the advection speed, which is smaller that the local fluid velocity, and thus $0<\mu<1$. 
In principle, $\mu$ should be a tensor for anisotropic particles but for simplicity we consider only particles which are weakly anisotropic and thus $\mu$ is assumed to remain a scalar~\footnote{Note that  the anisotropy of the mobility coefficient is much weaker in confined than in unbounded fluids due to the short range of hydrodynamic interactions in quasi 2D geometries.}. In addition to a change in their velocity, anisotropic particles experience hydrodynamic torques which favor an orientation along the local elongation axis of the flow. This classical hydrodynamic result, which can also be anticipated from symmetry arguments, leads to the so-called Jeffery's orbits~\cite{Jeffery1922}. As the flow  is irrotational (potential flow), the orientational dynamics reduces to
\begin{equation}
\partial_t\hat {\bf p}_i=\gamma \left( I-\hat {\bf p}_i \hat {\bf p}_i \right)\cdot{\bf E}({\bf R}_i)\cdot \hat {\bf p}_i,
\label{eq3}
\end{equation}
where ${\bf E}$ is the strain rate tensor, ${\bf E}=\frac{1}{2}\left[ \nabla {\bf V}+(\nabla {\bf V})^T \right]$,  and $\gamma\geq0$ is a rotational mobility coefficient which is non-zero for anisotropic particles and zero for axisymmetric bodies. 

\subsection{Long-range hydrodynamic interactions}
As a particle located at $\bf R(t)$ moves in the fluid, it induces a far-field velocity, denoted $\bf v(r-R)$, at  position $\bf r$.  A given particle $i$ responds to the flow induced by all the other particles in the crystal, and is therefore advected at  velocity  ${\bf V}({\bf R}_i)=\mu \sum_{j\neq i} {\bf v}({\bf R} _i-{\bf R}_j)$.  In this section we  provide a quantitative description of the far-field hydrodynamic coupling, $\bf v$, between the particles; for the sake of clarity, we  separate the case of passive and active particles.

\subsubsection{Hydrodynamic interactions between driven particles}
In the driven case, each particle of the crystal is subject to a constant external force, 
${\bf f}=f\hat{\bf x}$, which results in a far-field  perturbation which we denote ${\bf v}^{1}$ and is the Green's function of Eq.~\ref{potential}. In three-dimensional flows, the response to a force monopole is known as a Stokeslet, and decays spatially as $\sim1/r$. In our quasi-2D geometries,  solid walls act as momentum sinks and screen algebraically the Stokeslet contribution, which then decays as ${\bf v}^{1}\sim1/r^2$ and takes the functional form of a potential source dipole, as shown in~\cite{Liron:1976wz,Long:2001tn}. 
In addition,  the particles have a finite size and their advection by the surrounding fluid is hindered by the lubrication forces induced by the confining walls (even in the absence of external driving). Due to  incompressibility,   any relative motion with respect to the fluid results in another algebraic far field contribution, ${\bf v}^{2}$.  As shown, e.g. in~\cite{Beatus:2006hc}, ${\bf v}^{2}$ has also the form of a potential dipole  with the same spatial decay,  ${\bf v}^{2}\sim1/r^2$. 
(We note that in unbounded fluids, this potential contribution scales as $1/r^3$ and is thus subdominant with respect to the flow induced by a pointwise force, which decays as $1/r$). Therefore, in confined flows,  the two contributions  have the same form~\cite{Beatus:2006hc,Liron:1976wz}  and the overall flow disturbance,  ${\bf v}^{\rm d}={\bf v}^{1}+{\bf v}^{2}$,  takes the form of a $x$-dipole,
\begin{equation}
{\bf v}^{\rm d}({\bf r})=\frac{\sigma}{2\pi r^2}\left(2\hat{\bf r}\hat{\bf r}-I\right)\cdot\hat{\bf x},
\label{eq4}
\end{equation}
where $r=|\bf r|$ and  the dipole strength, $\sigma$, is the sum of the two contributions, $\sigma=A b^2Gf+Bb^2U_0$, where $A$ and $B$ are two dimensionless shape factors ($I$ is the identity tensor). The symmetry of the streamlines for this flow field are illustrated in Fig.~\ref{figdipole} (left).

\subsubsection{Hydrodynamic interactions between active swimmers}
\begin{figure}
\center
\resizebox{0.9\columnwidth}{!}{%
  \includegraphics{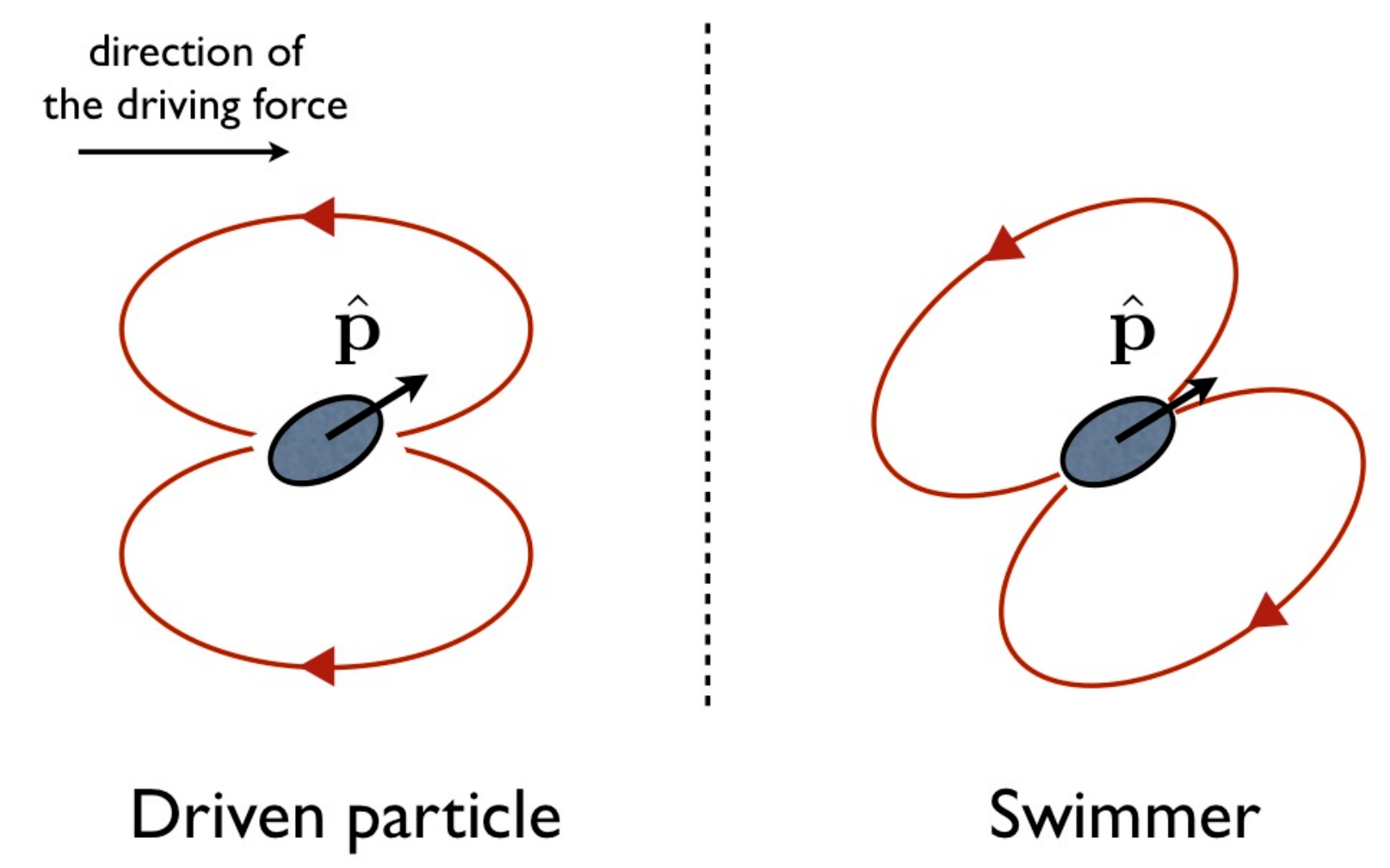}}
\caption{Sketch of the dipolar flow  field (potential source dipole) induced by  driven particles (left) and active swimmers (right).}
\label{figdipole}      
\end{figure}

By definition swimmers do not require an external force to propel themselves. The stress distribution on the surface of a self-propelled particle has thus, at least, the symmetry of a force dipole \cite{Lauga:2009p421}. The canonical theoretical setup used to describe (dilute) suspensions of swimmer is to consider an ensemble of such force-dipoles as all other multipolar contributions to the far field are subdominant in an unbounded fluid~\cite{Saintillan:2007jr,saintillan08,Baskaran:2009vu}.  However, as mentioned above, confinement  results in an algebraic screening of the hydrodynamic interactions. In the quasi-2D geometry at the center of our paper a force dipole decays spatially as $\sim1/r^3$, a contribution which is therefore subdominant compared to the $\sim1/r^2$ potential dipole arising from incompressibility (similarly to driven particles) \cite{Liron:1976wz}. 
For active swimmers, the far-field flow disturbance  has thus also the symmetry of a potential source dipole, the difference with the passive case being that the dipole direction is now the swimmer orientation (Fig.~\ref{figdipole}, right). For a swimmer orientated along $\hat {\bf p}$, we obtain a flow given by
\begin{equation}
{\bf v}^{\rm s}({\bf r},\hat {\bf p})=\frac{\sigma}{2\pi r^2}\left(2\hat{\bf r}\hat{\bf r}-I\right)\cdot\hat{\bf p},
\label{eq5}
\end{equation}
with the dipole strength $\sigma=Bb^2U_0$ ($B$ is the same shape factor as in Eq.~\ref{eq4}). We therefore see that, that in confined fluids, the usual distinction between pushers and pullers swimmers (contractile and extensile), which is at the heart of qualitatively different behaviors in unconfined fluids  \cite{Ramaswamy:2010bf,Kochreview}, is irrelevant. The magnitude and sign of the induced dipolar flow is solely set by that of the swimming speed, irrespective of the microscopic swimming mechanism.

\begin{figure*}
\center
\resizebox{0.97\textwidth}{!}{  \includegraphics{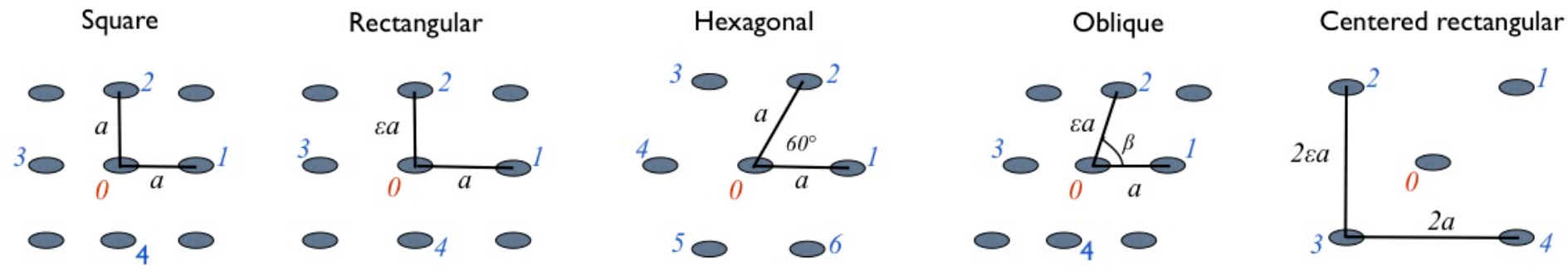}}
\caption{Geometry of the five planar Bravais lattices. Anisotropic cells are characterized by the ratio, $\epsilon$,  between the two cell dimensions. The angle $\beta$ is the tilt angle of the oblique and hexagonal cells. For each particle labeled ``0'' we also display and number all nearest neighbors.}
\label{fig2}      
\end{figure*}

In summary,  Eqs.~\ref{eq2}, \ref{eq3}, and either Eq.~\ref{eq4} (in the driven case) or \ref{eq5} (active case) fully  prescribe the dynamics of the discrete particle positions and orientations. As noted above, the main difference between active and passive particles concerns the orientation of the dipolar flow field: the orientation is slaved to the swimmer direction for active particles whereas it is constant and aligned along the x-direction for driven particles (the difference is further illustrated in Fig.~\ref{figdipole}). We will show in the next sections that this distinction markedly impacts the large-scale crystal dynamics. 

\section{Are hydrodynamic crystals stationary?}
When addressing the dynamics of an ordered phase, the first important question is whether this phase does correspond to a stationary state.  We focus here on the five planar Bravais crystals, which encompass all possible symmetries for bidimensional mono-atomic crystals (see Fig.~\ref{fig2}).

Let us first consider the case of  crystals composed of driven axisymmetric particles. The equations of motion reduce to
\begin{equation}
\partial_t{\bf R}_i=U_0\hat {\bf x}+\mu\sum_{j\neq i} {\bf v^{\rm d}}({\bf R}_i-{\bf R}_j),
\label{eq6}
\end{equation}
where the ${\bf R}_i$'s belong to one of the Bravais crystals from Fig.~\ref{fig2}. The lattice structure is conserved provided that  $\partial _t  ({\bf R}_i-{\bf R}_j)=0$  for all $i$ and $j$. It follows from Eq.~\ref{eq6} that $\partial _t  ({\bf R}_i-{\bf R}_j)=\mu\sum_{k\neq i} {\bf v^{\rm d}}({\bf R}_i-{\bf R}_k)-\mu\sum_{k\neq j} {\bf v^{\rm d}}({\bf R}_j-{\bf R}_k)$. By definition, all  crystals are invariant upon translation along $({\bf R}_i-{\bf R}_j)$, which  readily implies that the two sums are equal, and therefore that any  driven crystal  made of axisymmetric particles is stationary structure (i.e.~$\partial _t  ({\bf R}_i-{\bf R}_j)={\bf 0}$). 

To extend this result to driven crystals composed of anisotropic particles we first recall that all the Bravais lattices are invariant upon the parity transformation $\bf r\to-\bf r$. Moreover, as $\bf v^{\rm d}(r)$ is invariant upon this transformation whereas the sign of the gradient operator is reversed, we see that the strain rate tensor constructed from a superposition of potential source dipoles, Eq.~\ref{eq4}, transforms according to $\bf E(-r)=-\bf E(r)$. This implies that for any Bravais crystal, ${\bf E}({\bf R}_i)$ is identically zero anywhere  on the lattice. It follows from the  equation for the orientational dynamics, Eq.~\ref{eq3}, that $\partial_t \hat{\bf p}_i=0$ for all the particles. In conclusion, in the driven case, both the crystalline structure and the particle orientations remain stationary (in other words, the crystals are fixed points of the dynamical system).  

It is straightforward to generalize the above results to swimmer crystals. The equation of motion for the positions, Eq.~\ref{eq6}, is given by
\begin{equation}
\partial_t{\bf R}_i=U_0\hat {\bf p_i}+\mu\sum_{j\neq i} {\bf v^{\rm s}}({\bf R}_i-{\bf R}_j,\hat{\bf p}_j).
\label{eq7}
\end{equation}
Obviously, $\partial _t ({\bf R}_i-{\bf R}_j)$ cannot be zero if the $\hat{\bf p}_i$ are not all identical. Therefore, the crystal structure cannot be conserved if the initial orientation of the particles is not uniform -- in such cases  the crystal would ``melt''.  For uniform orientations,  say along $\hat {\bf x}$, Eqs.~\ref{eq6} and \ref{eq7} are identical, and so is the equation for the orientational dynamics since $v^{\rm s}({\bf r},\hat{\bf p}_i)=v^{\rm d}({\bf r})$. We are thus left with the same problem as in the driven case, which  implies that the structure of the crystals is conserved as long as the particles all swim along the same direction.

\section{Driven hydrodynamic crystals are marginally stable}

We start by  investigating in this section the linear stability of the five Bravais crystals with respect to perturbations in both the position and the orientation of the particles, with a special focus on the experimentally-relevant  square and hexagonal lattices.  Anticipating on our results, we note that the geometrical classification in term of the Bravais lattices might  not  necessarily be relevant to the dynamics of flowing crystals.  

In order to  proceed we  make use of two additional  assumptions. Firstly, we consider the case of  particles uniformly aligned along the $\hat {\bf x}$-axis prior to the perturbations, as depicted in Fig.~\ref{fig2}.   Secondly we assume that the driving force is aligned with one of the principal axes of the crystal. Our following study can be easily extended to a more general setup, but this would make the formula and the discussions much more tedious.

We  denote $\delta {\bf R}_i$ and $\delta \hat{{\bf p}}_i\sim \theta_i\hat {\bf y}$  the infinitesimal perturbations of  the particle positions and  orientations respectively, so that ${\bf R}_i\to {\bf R}_i+\delta {\bf R}_i$, and $\hat{{\bf p}}_i\to\hat{\bf x}+\theta_i\hat {\bf y}$. Using the property that ${\bf E}=0$ for dipoles organized into a Bravais lattice (as discussed in the previous section), and after some algebra, the linearization of the equations of motion, Eqs.~\ref{eq3} and \ref{eq6},  yields
\begin{equation}
\partial_t \delta{\bf R}_i=\mu {\sum_{j\neq i}}\left[\nabla{\bf v^{\rm d}}({\bf R}_{ij})\right]\cdot \delta{\bf R}_{ij}
\label{eq8},
\end{equation}
and,
\begin{equation}
\partial_t \theta_i=\frac{\gamma}{2}{\sum_{j\neq i}}\left(\nabla [\partial_{x}{v^{\rm d}_y}({\bf R}_{ij})+\partial_{y}{v^{\rm d}_x}({\bf R}_{ij})]\right)\cdot \delta{\bf R}_{ij},
\label{eq9}
\end{equation}
where ${\bf R}_{ij}={\bf R}_i-{\bf R}_j$, and $\delta {\bf R}_{ij}=\delta{\bf R}_i-\delta{\bf R}_j$. Eqs.~\ref{eq8} and \ref{eq9} dictate the dynamics of the elementary excitations in the frame where the unperturbed crystal is  stationary.  We note that the direction of the crystal translation is, in general, different from the driving direction. 

We now exploit the symmetries of the dipolar interactions. Inspecting the flow given by Eq.~\ref{eq4}, we  deduce that $\partial_x v_y^{\rm d}=\partial_y v_x^{\rm d}$, and $\partial_x v_x^{\rm d}=-\partial_y v_y^{\rm d}$. Using these relations, we look for plane waves solutions, $(\delta X_i, \delta Y_i, \theta_i)\equiv(\delta X, \delta Y, \theta)\exp(i\omega t-i{\bf q}\cdot {\bf R}_i)$. By doing so, we obtain a linear-stability system, which we write in  the generic form
 \begin{equation}
 \omega \left(
 \begin{array}{ cc }
    \delta X   \\
    \delta Y  \\
    \theta
  \end{array}\right)= \left(
 \begin{array}{ ccc}
    M_1& M_2 & 0    \\
     M_2 & -M_1 & 0  \\
  M_3 & M_4 & 0
  \end{array}\right) 
\left(
 \begin{array}{ cc }
    \delta X   \\
    \delta Y  \\
    \theta
  \end{array}\right)  
  \label{M},
 \end{equation}
where the coefficients of the stability matrix $\bf M$ are
\begin{eqnarray}
M_1&=&-i\mu\sum_{j\neq i}\left[1-\exp(i{\bf q}\cdot {\bf R}_{ij})\right]\partial_x v_x^{\rm d}({\bf R}_{ij})  \label{M1},\\
M_2&=&-i\mu\sum_{j\neq i}\left[1-\exp(i{\bf q}\cdot {\bf R}_{ij})\right]\partial_x v_y^{\rm d}({\bf R}_{ij})  \label{M2},\\
M_3&=&-i\gamma\sum_{j\neq i}\left[1-\exp(i{\bf q}\cdot {\bf R}_{ij})\right]\partial_{xx} v_y^{\rm d}({\bf R}_{ij})  \label{M3},\\
M_4&=&-i\gamma\sum_{j\neq i}\left[1-\exp(i{\bf q}\cdot {\bf R}_{ij})\right]\partial_{yx} v_y^{\rm d}({\bf R}_{ij})  \label{M4}.
\end{eqnarray}

We readily deduce from the matrix structure that a  perturbation in orientations only  would not induce any change in the crystal conformation. This is a direct consequence of the  dipolar coupling between the particles, $\bf v^{\rm d}$, which is only a function of the driving force direction and not of the particle orientation. On the contrary, perturbations in the position of a particle modify both position and orientation. In addition, perturbations in orientation only neither relax, grow or propagate. As the third column of the matrix $\bf M$ is always 0, this implies that it will always admit the eigenvalue $\omega_0=0$, associated to the pure-orientation  eigenmode $(0,0,1)$.

The other two eigenvalues of the $\bf M$ matrix are $\omega_\pm=\pm\sqrt{M_1^2+M_2^2}$. Exploiting again the fact that all  Bravais lattices are invariant upon parity transformation to write 
\begin{eqnarray}
M_1&=&-\frac{i\mu}{2}\sum_{j\neq i}\left[1-\exp(i{\bf q}\cdot {\bf R}_{ij})\right]\partial_x v_x^{\rm d}({\bf R}_{ij})\nonumber\\
&&-\frac{i\mu}{2}\sum_{j\neq i}\left[1-\exp(i{\bf q}\cdot {\bf R}_{ji})\right]\partial_x v_x^{\rm d}({\bf R}_{ji})\label{M1reel}.
\end{eqnarray}

By definition ${\bf R}_{ij}=-{\bf R}_{ji}$, and due to the dipolar symmetry of the hydrodynamic interaction, we have $\partial_xv_{x}^{\rm d}({\bf r})=-\partial_xv_{x}^{\rm d}(-\bf r)$. Using these two equalities in Eq.~\ref{M1reel}, we infer that  $M_1=-\mu\sum_{j\neq i}\sin({\bf q}\cdot {\bf R}_{ij})\partial_xv_x^{\rm d}({\bf R}_{ij})$, and therefore $M_1$ is a real number. Using the same method, and the identity $\partial_xv_{y}^{\rm d}({\bf r})=-\partial_xv_{y}^{\rm d}(-\bf r)$, one can show that $M_2$ is real as well. Therefore, for any  symmetry of the crystal, the pulsations of the plane waves, $\omega_\pm$, are real. In other words, for any Bravais lattice the crystal structure of driven particles is dynamically marginally stable.

Notably, the dipole strength $\sigma$ can be eliminated from the equations of motion Eqs.~\ref{eq8} and \ref{eq9}  by rescaling the timescale. Therefore, the linear stability  of the monocrystals is a purely geometrical problem. The corresponding eigenmodes do not depend on the translational speed $U_0$, but only on the orientation, and on the symmetries of the lattice.

Interestingly, we see that  phonons propagate with the pulsations $\omega_{\pm}$, despite the fact that particles have no inertia and that no potential forces couple the particle displacements.  This seemingly counterintuitive result generalizes the experimental observations made by Beatus and coworkers in \cite{Beatus:2006hc} where they revealed that sound modes propagate along 1D-droplet crystals flowing in quasi-2D microchannels. These results are, importantly, specific to the quasi-2D geometry, which is relevant for numerous microfluidic and thin films applications. In unbounded fluids, the change in the symmetry of the hydrodynamic interactions results in the destabilization of the crystal structure as shown theoretically and experimentally \cite{CROWLEY:1976tt}.

Below, we derive the dispersion relation for each of the five Bravais crystals, with a special attention given to the case of  square and hexagonal lattices

\subsection{Square lattice}
\begin{figure}
\center
\resizebox{0.8\columnwidth}{!}{\includegraphics{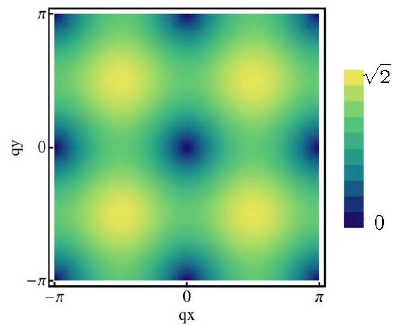}}
\caption{Normalized dispersion relation for the square lattice plotted from Eq.~\ref{eq16} ($\omega_+$ only), for ${2\mu \sigma}/{\pi a^{2}}=1$, and $a=1$.}
\label{fig3}      
\end{figure}

In order to compute the coefficients of the $\bf M$-matrix analytically we now make a  nearest neighbor approximation. In a similar context, this approximation has proven to yield qualitatively correct results for unbounded fluids \cite{CROWLEY:1976tt}. 
We introduce a reference particle labelled as $0$. The four nearest neighbors in the square crystal are labeled as 1, 2, 3, and 4 (Fig.~\ref{fig2}). In this geometry, we easily compute the coefficient of the $\bf M$ matrix as
 \begin{equation}
{\bf M}=\frac{2\mu \sigma}{\pi a^{3}} \left(
 \begin{array}{ c c c}
    \sin(q_{x}a) & -\sin(q_{y}a) & 0    \\
     -\sin(q_{y}a) & -\sin(q_{x}a) & 0  \\
    0 & -\frac{3\gamma i}{\mu a}[\cos(q_{x}a)+\cos(q_{y}a)-2] & 0
  \end{array}\right).
\label{eq15}
  \end{equation}
 \normalsize 

The dispersion relation of the infinitesimal excitations can be deduced by diagonalizing $\bf M$. The three eigenvalues are $\omega_0=0$, and
\begin{equation}
\omega_{\pm}=\pm \frac{2\mu\sigma}{\pi a^{3}}\sqrt{\sin^2(q_{x}a)+\sin^2(q_{y}a)}.
\label{eq16}
\end{equation}
This dispersion relation is  plotted in Fig.~\ref{fig3}.

To gain insight into the propagating modes, we focus on the large-scale (long wavelength) response of the crystals. Expanding Eq.~\ref{eq15} at leading order in the wave vector amplitude for $q \to 0$, we find
\begin{equation}
{\bf M}=\frac{2\mu \sigma}{\pi a^{2}} \left(
 \begin{array}{ ccc}
    q_{x} & -q_{y} & 0    \\
     -q_{y} & -q_{x} & 0  \\
   0 & 0 & 0
  \end{array}\right) +{\cal O}(q^2).
  \label{eq17}
\end{equation}

We  notice that in this small-$q$ limit, the orientation and the position degrees of freedom are totally decoupled.  The three eigenvalues are $\omega_0=0$, and $\omega_\pm=\pm2\mu\sigma q/(\pi a^2)$. The two non-trivial modes are non-dispersive and propagate with a constant ``sound-velocity'' $c_\pm=\pm2\mu\sigma/(\pi a^{2})$, which  increases  with the magnitude of the hydrodynamic coupling.

\begin{figure}
\center
\resizebox{0.6\columnwidth}{!}{ \includegraphics{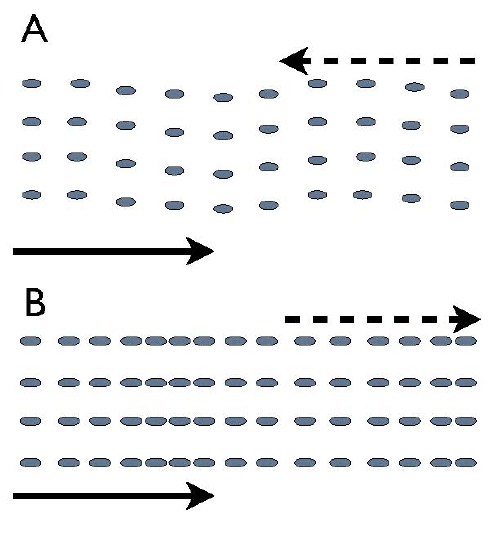}}
\caption{Sketch of the propagative eigenmodes in a square lattice for $q_y=0$. The full line corresponds to the direction of the driving, the dotted line indicates the direction of the wave propagation. A: Shear modes; B: Compression modes. The particle orientations are not affected by the perturbation.}
\label{fig4}      
\end{figure}

To understand physically how the excitations propagate, we focus on two specific cases. Let us first consider longitudinal perturbations,  ${\bf q}=q\hat{\bf x}$. The mode 
$\omega_{-}$ is here associated with the eigenvector $(0,1,0)$. It corresponds to shear waves which propagate in the direction opposite to the driving, as illustrated in  Fig.~\ref{fig4}A. 
The second sound mode ($\omega_+$)  corresponds to compression waves along the $x$-axis propagating in the driving direction, see Fig.~\ref{fig4}B. The corresponding eigenvector is $(1,0,0)$. 

For excitations propagating in the direction transverse to the driving, ${\bf q}=q\hat{\bf y}$, the eigenmodes couple the displacements along the two principal axes of the crystal.
 The mode $\omega_{-}$ is associated with the eigenvector $(1,1,0)$. It corresponds to the superposition of a compression mode in the $\hat{\bf y}$ direction, in phase with a shear in the $\hat{\bf x}$ direction. The second mode ($\omega_+$), with eigenvector $(-1,1,0)$,  is a combination of a dilation in the $\hat{\bf y}$ direction, which propagate in antiphase with a shear wave in the $\hat{\bf x}$ direction.

To close, we note that  the dispersion relation of the phonons remains unchanged  if the driving force is not aligned with one of the principal axes of the crystal, although in that case the form of the eigenmodes is more complex.

 \subsection{Hexagonal lattice}  
We  now consider the case of the hexagonal lattice. The main technical difference with the square lattice is that the reference particle $0$ has now six nearest neighbors, see Fig.~\ref{fig2}. Repeating the same procedure as above, we compute the coefficients of the stability matrix and obtain
\begin{equation}
{\bf M}=\frac{2\mu \sigma}{\pi a^{3}} \left(
 \begin{array}{ c c c}
    M'_1  & 0 & 0    \\
   0  & -M'_1 & 0  \\
   M'_3  &  M'_4 & 0
  \end{array}\right)
  \label{eq18},
  \end{equation}
where
 \begin{equation}
\left\{
\begin{array}{lll}
M'_1=\sin(q_{x}a)-2\sin(\frac{q_{x}a}{2})\cos(\frac{q_{y}\sqrt{3}a}{2}), \\
M'_3= \frac{3\gamma\sqrt{3}i}{\mu a}\sin(\frac{q_{x}a}{2})\sin(\frac{q_{y}\sqrt{3}a}{2}),\\
M'_4=-\frac{3\gamma i}{\mu a}[\cos(q_{x}a)-\cos(\frac{q_{x}a}{2})\cos(\frac{q_{y}\sqrt{3}a}{2})].
\end{array}\right.
\label{eq19}
\end{equation} 
Notably,  the upper-left $2\times2$ sub-bloc of $\bf M$  is diagonal. As a consequence,   an excitation of the position along one direction ($x$ or $y$) induces no net displacement in the transverse  direction. 
\begin{figure}
\center
\resizebox{0.8\columnwidth}{!}{  \includegraphics{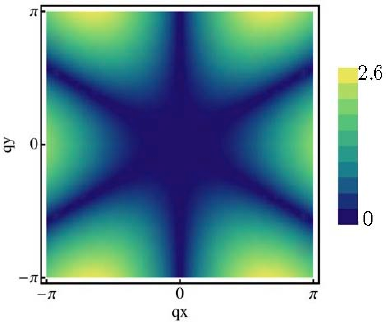}}
\caption{Normalized dispersion relation for the hexagonal lattice plotted from Eq.~\ref{eq20} ($\omega_+$ only), for ${2\mu \sigma}/{\pi a^{2}}=1$, and $a=1$.}
\label{fig7}      
\end{figure}
The dispersion relations, plotted in Fig.~\ref{fig7}, is given by
\begin{equation} 
\omega_{\pm}=\pm \frac{2\mu\sigma}{\pi a^{3}}\left|\sin(q_{x}a)-2\sin(\frac{q_{x}a}{2})\cos(\frac{q_{y}\sqrt{3}a}{2})\right|.
\label{eq20}
\end{equation}

We see from Eq.~\ref{eq20} that there exist two specific orientations of the wavevectors for which no excitation propagates. For perturbations making angles equal to  $\pi/6$, (resp.~$\pi/2$)  with the $x$-axis,  we obtain $M'_1=0$ (resp.~$M'_1=M'_3=0$); in these cases, the matrix is not diagonalizable and the only solutions of Eq.~\ref{eq18} is $\omega=0$. Phonons therefore do not propagate in those two directions.

To illustrate the difference in the dynamics between the hexagonal and the square crystals we consider the behavior in the long wavelength limit. Expanding Eq.~\ref{eq18} at leading order in $q$, we obtain
\begin{equation}
{\bf M}=\frac{2\mu\sigma}{\pi } \left(
 \begin{array}{ c c c}
    -\frac{1}{8}q_{x}^{3}+\frac{3}{8}q_{y}^{2}q_{x} &  0 & 0    \\
     0 &  \frac{1}{8}q_{x}^{3}-\frac{3}{8}q_{y}^{2}q_{x} & 0  \\
   \frac{9\gamma}{4\mu a^2}i q_{x}q_{y} & \frac{9\gamma}{8\mu a^2}i(q_{x}^2-q_{y}^2) & 0
  \end{array}\right),
   \end{equation}  
and the eigenvalues takes the form
$\omega_{\pm}=\pm \frac{\mu \sigma}{4\pi }|3q_xq_y^2-q_x^3|$. We infer from this formula that hydrodynamic crystals having an hexagonal symmetry are ``softer'' than square crystals. When $q\to0$, the speed of sound goes to 0 as $q^2$ and even  the large-wavelength phonons are dispersive. Furthermore, the sound modes couple the displacements and the orientation of the particles. 

To convey a more intuitive picture, we again focus on two specific directions of propagation. We first consider longitudinal perturbations along the first principal axis of the crystal, ${\bf q}=q\hat {\bf x}$. As above, we find that one of the eigenvector correspond to a pure compression along the $x$-axis, $(1,0,0)$. The second eigenvector, $(0,-iq/9,1)$, mixes  shear  and orientational waves (bending modes) in quadrature, and depends explicitly on $q$. Such a coupling was not observed for the square lattice. A second simple case concern the excitations propagating along the second principal direction, namely ${\bf q}=\cos(\pi/3)\hat {\bf x}+\sin(\pi/3)\hat {\bf y}$. Here, the two eigenmodes mix the particle displacements (in only one of the two directions, since $x$ and $y$ cannot couple) and their orientation. They are given by $(0,-2iq/9,1)$ and $(-2i q/(9\sqrt{3}),0,1)$ and correspond to $\omega_-$ and $\omega_+$ respectively.

 \subsection{Rectangular,oblique, and face-centered lattices}
\begin{figure}
\center
\resizebox{1\columnwidth}{!}{  \includegraphics{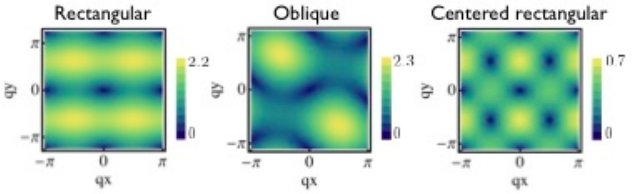}}
\caption{Normalized dispersion relations for the rectangular, oblique and centered rectangular lattices  ($\omega_+$ only), for ${2\mu \sigma}/{\pi a^{2}}=1$, and $a=1$. Rectangular: $\epsilon=0.8$. Oblique: $\epsilon=0.9$ and $\beta=(\pi/2)-0.2$. Centered rectangular: $\epsilon=0.9$.}
\label{fig8}      
\end{figure}

The lattice geometries corresponding to the rectangular, the oblique and the face-centered lattices are shown in Fig.~\ref{fig2}. To derive the eigenmodes we restrict our analysis to calculations with four nearest neighbors, an assumption which restrains the number of crystals for which our calculations are correct (weakly anisotropic and weakly tilted lattices only, as sketched in Fig.~\ref{fig2}). It is straightforward to proceed mathematically as in the two previous cases and derive the two sound modes, $\omega_\pm$,  propagating. The results for the dispersion relation are plotted in Fig.~\ref{fig8}. In the small $q$-limit, these phonons always propagate  in a dispersive manner.
As $q$ goes to zero, the sound velocities reach a constant value  which depends on the orientation of the propagation due to the crystal anisotropy.

 \subsection{Response of driven hydrodynamic crystals to finite amplitude  perturbations}
Before closing this section, we make a final remark regarding the stability of all the five Bravais crystals with respect to finite amplitude perturbations. We start by a  simple observation on the relationship between the various lattices. A rectangular lattice corresponds to a square lattice transformed upon a finite homogeneous stretching. An oblique lattice is obtained by stretching and shearing a square lattice. The hexagonal lattice is an oblique lattice with a tilt angle of $\pi/3$. Finally, a face-centered lattice is obtained from a rectangular lattice by applying a shear modulated at the highest possible wavelength ($q=2\pi/a$).  As all these structures are stationary, and marginally stables at the linear level, we can deduce that any finite amplitude deformation corresponding to an homogeneous shear, or stretch, of the crystal would also be a marginal perturbation: their growth rate would be zero. The same conclusion also holds for rectangular crystals deformed by the specific high-$q$ shear that would transform them into a face-centered lattice.

\section{Hydrodynamic stability of active crystals}
We now move on to investigate the linear stability of active swimmer crystals. To do so, we use the same theoretical framework as in the previous section. The swimmers self-propel  along one of the principal axes of the crystals.  We also recall that in this active case, the swimming direction is slaved to the particle shape and so is the dipolar flow (Fig.~\ref{figdipole}).
Following the same strategy as in the case of driven particles, we first establish the linearized equations of motion. Combining Eqs.~\ref{eq3}, \ref{eq5} and \ref{eq7} we obtain
\begin{eqnarray}
\partial_{t}\delta{\bf R}_i&=&U_{0}\theta_i\mathbf{e}_{y}\label{eq23}\\
&+&\mu {\sum_{j\neq i}}\left(\left[\nabla{\bf v^{\rm s}}({\bf R}_{ij},{\hat {\bf x}})\right]\cdot \delta{\bf R}_{ij}+\left[\partial_{\theta}{\bf v^{\rm s}}({\bf R}_{ij},{\hat {\bf x}})\right] \theta_j\right)
\nonumber,
\end{eqnarray}
and
\begin{eqnarray}
\partial_{t}\theta_i=\gamma{\sum_{j\neq i}}&&[\mathbf{\nabla}\left [\partial_{x} {v}_{y}^{\rm s}({\bf R}_{ij},{\hat {\bf x}})]\cdot \delta{\bf R}_{ij}\right .\label{eq24}\\
&+&\partial_{\theta}E^{\rm s}_{yx}({\bf R}_{ij},{\hat{\bf p}}_j=\hat{\bf x})\theta_j]\nonumber,
\end{eqnarray}
where $E^{\rm s}_{yx}$ is the $(y,x)$ component of the strain rate tensor associated with the dipolar perturbation induced by the swimmer located at $j$, namely ${\bf v}^{\rm s}({\bf R}_{ij},{\hat{\bf p}}_j=\hat{\bf x})$. 

Two important remarks can be made at this point. First we see that  the stability equations now depend explicitly on the swimming speed of the particles. In addition, contrary to driven lattices, the stability of the swimmer crystals depends on the particle shape through $\gamma$. Therefore, we discuss below  isotropic and anisotropic swimmers separately.

\subsection{Isotropic swimmers}
Isotropic particles correspond to $\gamma=0$. Their dynamic equation are significantly simplified as Eq.~\ref{eq24} is now trivial and the swimmer orientation remains constant. Since the flow is irrotational, no hydrodynamic torque (from vorticity) is present to modify the orientations of the particles. 
 As in the previous section on driven suspensions, we look for plane waves solutions, from which we infer the form of the stability matrix $\bf M$. This matrix is analogous to the one defined in Eq.~\ref{M} but takes  here a slightly different structure
\begin{equation}
{\bf M}=\left(
 \begin{array}{ c c c}
    M_1  & M_2 & M_5    \\
   M_2  & -M_1 & M_6  \\
   0  &  0 & 0
  \end{array}\right),
  \label{eq25}
\end{equation}
where $M_1$ and $M_2$ are given by Eq.~\ref{M1}, and Eq.~\ref{M2} respectively. The two new coefficients are
\begin{eqnarray}
M_5&=&-i\mu\sum_{j\neq i}\left[1-\exp(i\bf q\cdot {\bf R_{ij}})\right]\partial_\theta v_x^{\rm s}({\bf R_{ij}},{\hat {\bf x}}) \label{M5},\\
M_6&=&-iU_0-i\mu\sum_{j\neq i}\left[1-\exp(i\bf q\cdot {\bf R_{ij}})\right]\partial_\theta v_y^{\rm s}({\bf R_{ij}},{\hat {\bf x}}).\,\,\,\,\,\, \label{M6}
\end{eqnarray}

Independently of the crystal symmetry, we see that the eigenvalues of the above matrix are identical to the one we found for driven crystals, $\omega_0=0$, and $\omega_\pm=\pm\sqrt{M_1^2+M_2^2}$. Therefore,  crystals composed of isotropic swimmers are marginally stable and phonons propagate with the same dispersion relations as in driven lattices, albeit with different eigenmodes.

\subsection{Anisotropic swimmers}
We now explore the richer phenomenology arising from swimmer anisotropy. Generic results cannot be established in a framework as general as in the isotropic case. We  proceed with the calculation under the nearest neighbors approximation, and deal with the five Bravais lattices separately.

\subsubsection{Square lattice}
To establish the linear stability of the square crystal we compute all the coefficients of the $\bf M$ matrix using Eqs.~\ref{eq23} and~\ref{eq24} and obtain
\begin{equation}
{\bf M}=\frac{2\mu\sigma}{\pi a^3}\left(
 \begin{array}{ c c c}
    M'_1  & M'_2 & 0    \\
   M'_2  & -M'_1 & M'_6  \\
   0  & M'_4 & M'_7
  \end{array}\right),
\end{equation}
with
\begin{eqnarray}
M'_1&=&\sin(q_xa),\\
M'_2&=&\sin(q_ya),\\
M'_4&=&-3i\frac{\gamma}{\mu a}\left[\cos(q_xa)+\cos(q_ya)-2\right],\\
M'_6&=&-ia \left( p+ \frac{1}{2} \left[\cos(q_ya)-\cos(q_xa)\right]\right),\\
M'_7&=&\frac{\gamma}{\mu}\sin(q_xa),
\end{eqnarray}
where we introduced the dimensionless number
\begin{equation}
p\equiv \frac{\pi U_0a^2}{2\mu\sigma}\cdot
\end{equation}

\begin{figure}
\center
\resizebox{0.6\columnwidth}{!}{%
  \includegraphics{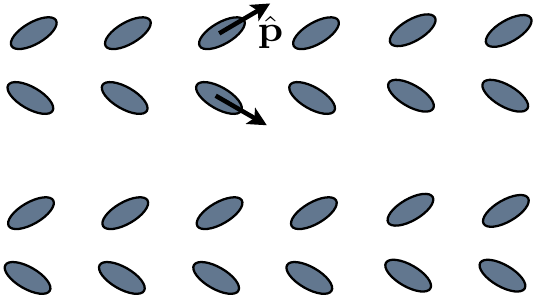}
}
\caption{Unstable position/orientation mode for a square lattice of active particles if the volume fraction is high enough ($p < 1$). The mode is a compression along the $y$-direction out of phase with a splay perturbation of the particles orientation and leads to the formation of short wavelength bands.}
\label{fig10}      
\end{figure}

Unlike the driven case, the stability matrix for active particles is not characterized solely by the  geometry of the lattice. The parameter $p$ quantifies the relative magnitude of the swimming speed and the dipolar advection velocity induced by a neighboring particle. Recall that $\sigma$ is itself a function of $U_0$, and of the particle shape, and $\sigma\equiv BU_0 b^2$, where $B$ is a shape factor of order $1$. Therefore, $p$ scales as $p\sim \mu^{-1} (a/b)^2$. Large values of $p$ correspond to the dilute limit, $a \gg b$, in which our far field approach is expected to be quantitatively correct. Small values of $p$ correspond to a dense crystal, for which our model should capture the essential physical features. The presence of $p$ in the matrix $\bf M$ means that the crystal stability now strongly depends on the particle volume fraction. 

As even in the large-$p$ limit  the eigenvalues of $\bf M$ take a quite complex form, we proceed to consider  the short and long wavelength excitations separately. In the limit $q\to0$, the matrix $\bf M$ has again three real eigenvalues, corresponding to  three propagating modes  with frequencies $\omega_0=2\gamma\sigma q_x/(\pi a^2)$, and $\omega_\pm=\pm2\mu\sigma q/(\pi a^2)$. The mode $\omega_0$ is a combination of phonons and orientation waves, whereas $\omega_\pm$ are the phonon modes we found for driven crystals (Fig.~\ref{fig4}).

In the high-$q$ limit (small wavelengths) the phenomenology is markedly different. For waves vectors of the edge of the Brillouin zone,  $q_x=0$ and $q_y=\pi/a$, we find $\omega_0=0$ as well as two non-trivial modes,  $\omega_\pm=\pm\frac{2\sigma}{\pi a^3}\sqrt{6\gamma\mu(p-1)}$. Importantly, $\omega_\pm$ are either real or pure imaginary numbers depending on the magnitude of $p$. In principle, $p>1$ for dilute crystals, and therefore the modes $\omega_\pm$ correspond again to a combination of phonons and orientation waves. However, we can expect our results to hold at a qualitative level for more concentrated systems, for which $p<1$. In such a case, the hydrodynamic coupling destroys the square crystal structure. Specifically, the $\omega_-$ mode is unstable. It correspond to the eigenvector $(\delta X,\delta Y, \theta)=(0,-i\frac{\pi a^3}{6\gamma\sigma}\omega_-,1)$, which combines a compression along the $y$-axis  and  splay distortions of the particle orientation. In this strong hydrodynamic coupling limit, the square crystal evolves to form short wavelength bands aligned with the average swimming direction, as sketched in Fig.~\ref{fig10}.

\begin{figure}
\center
\resizebox{0.78\columnwidth}{!}{ \includegraphics{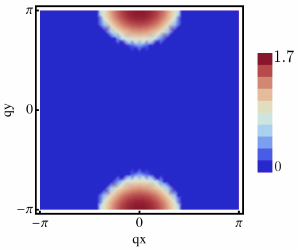}}
\caption{Normalized growth rate ($i\omega_-$)  of the $(q_x,q_y)$ modes of the square lattice of active particles for $p=1/2$. The parameters are  ${2\mu \sigma}/{\pi a^{2}}=1$, $a=1$ and $\gamma/\mu=1$.}
\label{fig11}      
\end{figure}

At second order in $q \rightarrow 0$ and given that $p$ is small enough, the eigenvalues $\omega_{\pm}$ have a non-zero imaginary part which scales as $q^2$. These eigenvalues correspond to the roots of a 3rd-order polynomial, which has no analytical solution. Therefore, we proceed to a numerical investigation of the short wavelength dynamics of the crystal. We compute numerically the eigenvalues of the matrix $\bf M$ for all $q$'s and  $0<p<3$. We find that the square crystals are indeed always unstable for $p<1$. In addition the wavenumbers $q_x=0$ and $q_y=\pi/a$ correspond to the most unstable mode as shown in Fig.~\ref{fig11} for $p=1/2$.

\subsubsection{Hexagonal lattice}

When the lattice has hexagonal symmetry, the structure of $\bf M$ is somewhat  simplified and we obtain in this case
\begin{equation}
{\bf M}=\frac{2\mu\sigma}{\pi a^3}\left(
 \begin{array}{ c c c}
    M'_1  & 0 & \frac{\mu a^2}{6\gamma}M'_3    \\
   0  & -M'_1 & -iap-\frac{\mu a^2}{2\gamma}M'_4  \\
   M'_3  & M'_4 & \frac{\gamma}{\mu}M'_1
  \end{array}\right),
\end{equation}
with,
\begin{eqnarray}
M'_1&=&\sin(q_xa)-2\sin(q_xa/2)\cos(\sqrt{3}q_ya/2),\\
M'_3&=&i\frac{3\gamma\sqrt{3}}{\mu a}\sin(q_xa/2)\sin(\sqrt{3}q_ya/2),\\
M'_4&=&3i\frac{\gamma}{\mu a}\left[\cos(q_xa/2)\cos(\sqrt{3}q_ya/2)-\cos(q_xa)\right].\,\,\,\,\,
\end{eqnarray}

\begin{figure}[t]
\center 
\resizebox{0.75\columnwidth}{!}{  \includegraphics{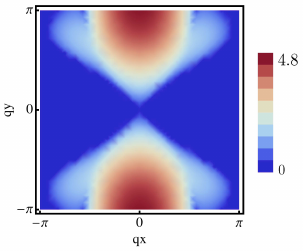}}
\caption{Normalized growth rate ($i\omega_-$)  of the $(q_x,q_y)$ modes of the hexagonal lattice of active particles for $p=5$. The parameters are ${2\mu \sigma}/{\pi a^{2}}=1$, $a=1$ and $\gamma/\mu=1$.}
\label{fig12}      
\end{figure}

Similarly to the square lattice, $\bf M$ does depend on the relative magnitude of the hydrodynamic coupling through $p$.  The general form of the eigenvalues is too complex to yield an intuitive picture. However, the salient features correspond to small wave vectors. In this limit $q\to0$, the  eigenvalues of $\bf M$ are $\omega_0=0$, and $\omega_\pm=\pm\frac{\sigma}{\pi a^2}\sqrt{\frac{9}{2}\gamma\mu p(q_x^2-q_y^2)}$. For all values of  $p$, there exists therefore an infinite number of unstable modes growing at a rate $|\omega_\pm|$. They correspond to perturbations in the position of the particle propagating along a direction making an angle comprised between $[\pm \pi/4,\pm3\pi/4]$ with respect to $x$. This behavior is illustrated in Fig.~\ref{fig12}, where we show the variations of the imaginary part of $\omega_\pm$ in the $(q_x,q_y)$ plane for $p=5$. We note that the most unstable mode is again a combination of  compression along the $y$-axis and splay-like instability of the particle orientation. 

 \subsubsection{Rectangular crystals}
The behavior for the rectangular lattices is very similar to what we found for square lattice (within the limits of the nearest neighbor approximation). These crystals are all stable for long wavelengths but  can display short wavelength instabilities. Denoting $\epsilon$ the aspect ratio of the lattice cell (Fig.~\ref{fig2}) a numerical diagonalization of the stability matrix reveals that again the most unstable mode lies on the edge of the Brillouin zone in the $y$ direction. The associated eigenvalue is $\omega_\pm=\frac{\sigma}{\pi(\epsilon a)^3}\sqrt{3\mu \gamma\left[(2p-1)\epsilon^2-1\right]}$. Hence, there exists a  critical value $p_c=\frac{1}{2}(1+\epsilon^{-2})$ such that the crystal destabilizes for $p<p_c$; note that $p_c$  is a decreasing function of the aspect ratio which plateaus at $p=1/2$. Dilute crystals corresponding to high $p$-values are stable and display phonons modes. Note that, similarly to our observation on the rectangular driven lattices, the latter result implies that dilute swimmer crystals with a rectangular symmetry are marginally stable with respect to finite amplitude stretch deformations.

\subsubsection{Oblique and face-centered lattices}
The dynamics of active crystals having oblique, or face centered symmetries are much more complex. We here briefly highlight some interesting large scale properties, and comment on the stability of these structures. 

The stability matrix of the oblique crystals take a simple for the global modes only, $q=0$, yet it reveals an original dynamics. Indeed for $q=0$ we get 
\begin{equation}
{\bf M}=\frac{2\mu\sigma}{\pi a^3}
\left(
\begin{array}{ccc}
 0 & 0 & -\frac{i a \cos (\beta ) \sin (\beta )}{\epsilon ^2} \\
 0 & 0 & -i p a+\frac{i \cos (2 \beta ) a}{2 \epsilon ^2}+\frac{i a}{2} \\
 \frac{3\gamma i \sin (4 \beta )}{\mu a \epsilon ^4} & 0 & 0
\end{array}
\right),
\end{equation}
where $\beta$ is the inclination of the lattice cells and $\epsilon$ their aspect ratio (Fig.~\ref{fig2}). Recall that the nearest neighbor scheme restrains our analysis to weakly tilted and weakly anisotropic lattices. 
Beyond the $\omega_0=0$ mode, the other two eigenvalues are nonzero, and we obtain 
\begin{equation}
\omega_\pm=\frac{\sigma}{\pi (\epsilon a)^3}\sqrt{3\mu \gamma[\cos(2\beta)-\cos(6\beta)]}.
\end{equation}

For weakly tilted lattices $\beta\approx\pi/2$, so that the frequencies are purely imaginary, and thus  $q=0$ modes are unstable. Note that this result does not contradict the stationarity of the structure. Indeed, the orientation field is here unstable, thereby inducing a coupled translation of the lattice, as swimmers rotate.

Conversely, in the small-$q$ limit, the face-centered lattices are marginally stable for any  amplitude of the hydrodynamic coupling, and phonons and orientation waves propagate in a non dispersive manner. For small wavelengths however, and looking specifically at the combination $(q_x=0,q_y=\pi/\epsilon a)$, we see that the eigenmodes corresponding to compression  along the $y$-direction coupled to distortions of the orientation grow exponentially at a rate $\frac{2\sigma}{\pi(\epsilon a)^3}\sqrt{3p\gamma\mu}$.  This last results implies that face-centered  swimmer lattices are unstable for any amplitude of the hydrodynamic coupling.

\section{Conclusion}

In this paper we considered theoretically the dynamics and stability of both driven and active crystals. With a geometry of elongated particles under confinement we derived the dynamical system quantifying the time-evolution of the  particle positions and orientations and showed that all five planar Bravais lattices are stationary solutions of the equations of motion.  In the case of particles passively driven by an external force we formally demonstrated that all five lattices are always marginally stable. The phonons modes do not depend on the magnitude of the driving force  but solely on the orientation and on the symmetries of the lattices. We detailed the spatial structure of the eigenmodes in the square and hexagonal geometry. 

In the separate  case where the particles are actively self-propelling we showed that the stability of the particle positions and orientations depends not only on the symmetry of the crystals but also on the perturbation wavelengths and  the volume fraction of the crystal. We obtained that the square and rectangular lattices are linearly unstable at short wavelengths provided the volume fraction of the crystals is high enough. As a difference, hexagonal, oblique, and face-centered crystals are always unstable. 

The results of our work can be compared with past theoretical studies. In the driven case,  planar crystalline arrangements,  were shown to be hydrodynamically unstable in a three-dimensional fluid at long wavelengths \cite{CROWLEY:1976tt}. The results in our paper demonstrate that confinement of the crystals, which algebraically screens hydrodynamic interactions between the particles, leads to a qualitatively different behaviors and all lattices solely support phonon modes.

In the active case, previous work demonstrated the presence of long wavelength   instabilities in orientation, density and stress (see \cite{Ramaswamy:2010bf,Kochreview} and references therein). In this past work,  aligned suspensions for both pusher and puller swimmers were shown to be unstable in the dilute regime, and so are  isotropic suspensions of pushers  \cite{SaintillanShelley2008} whereas isotropic puller suspensions, which are linearly stable at zero volume fraction, were shown numerically to be unstable  at high volume fraction \cite{Art}. In our paper, again because of hydrodynamic screening, the stability characteristics of confined active crystals were found to be independent of the pusher vs.~puller nature of the self-propelled particle - the only flow singularity  dictating hydrodynamic interactions in this case is the potential flow dipole whose sign is set by the swimming direction only.

\section*{Acknowledgements}
This work was funded in part by the NSF (grant 0746285 to E.L.). We acknowledge support from Paris Emergence research program, and C'Nano Idf.

\bibliographystyle{epj}
\bibliography{BIBTEXXTALS}

\end{document}